\documentclass[12pt]{article}

\usepackage[utf8]{inputenc} %unicode support
\usepackage{algorithm}
\usepackage{algpseudocode}
\usepackage{float}
\usepackage{url}
\algdef{SE}[DOWHILE]{Do}{doWhile}{\algorithmicdo}[1]{\algorithmicwhile\ #1}
\usepackage{graphicx}

\title{Lock-free de Bruijn graph}
\author{Daniel Górniak\footnote{daniel.gorniak@op.pl} \and Robert Nowak\footnote{ORCID 0000-0001-7248-6888}\footnote{Correspondence: robert.nowak@pw.edu.pl}}
\date{Institute of~Computer Science, Warsaw University of~Technology, Warsaw, 00-665, Poland}
\begin{document}

\maketitle

\begin{abstract} % abstract
  \begin{minipage}{\textwidth}
  \paragraph{Background.} De Bruijn graph is one of the most important data structures used in de-novo genome assembly algorithms, especially for NGS data.
There is a growing need for parallel data structures and algorithms due to the increasing number of cores in modern computers.
The assembly task is an indispensable step in sequencing genomes of new organisms and studying structural genomic changes. In recent years, the dynamic development of next-generation sequencing (NGS) methods raises hopes for making whole-genome sequencing a fast and reliable tool used, for example, in medical diagnostics. However, this is hampered by the slowness and computational requirements of the current processing algorithms, which raises the need to develop more efficient algorithms. One possible approach, still little explored, is the use of quantum computing.

\paragraph{Results.} We created the lock-free version of the de Bruijn graph, as well as a lock-free algorithm to build such graph from reads.
Our algorithm and data structures are developed to use parallel threads of execution and do not use mutexes or other locking mechanisms,
instead, we used only compare-and-swap instruction and other atomic operations.
It makes our algorithm very fast and efficiently scaling.

\paragraph{Conclusions.} The presented article depicts the new lock-free de Bruijn graph data structure with a graph build algorithm.
We developed a C++ library and tested its performance to depict its high speed and scalability compared to other available tools.
  \end{minipage}
\end{abstract}

%%%%%%%%%%%%%%%%%%%%%%%%%%%%%%%%%%%%%%%%%%%%%%
%%                                          %%
%% The keywords begin here                  %%
%%                                          %%
%% Put each keyword in separate \kwd{}.     %%
%%                                          %%
%%%%%%%%%%%%%%%%%%%%%%%%%%%%%%%%%%%%%%%%%%%%%%


\section{Background}

Sequencing costs have dropped significantly in the last few years, increasing the number of sequencing projects.
Assembly is a part of sequencing projects, and it is a challenging computational task where the consensus is built from sequencers output, e.g. set of reads.
Although many genome assemblers have been presented, there are still significant challenges for de-novo genome assembly that our algorithms address. The most important is reducing the huge computational resource consumption.

De Bruijn graphs and its extensions (A-Bruijn, paired graph, etc.) are commonly used in de-novo assembly
due to its high usability for NGS data. Such graph uses substrings $s_{0}s_{1}\dots{}s_{k-1}$ of length $k$ of reads as edges,
where $s_{0}\dots{}s_{k-2}$ is sequence of source vertex, $s_{1}\dots{}s_{k-1}$ is sequence of target vertex.
The assembly output is an eulerian path over the graph.

\begin{figure}[!htb]%figure1
  \caption{A-Bruijn, $4$ dimension ($k=4$) graph for $ACGACGACGC$. Picture shows sub-sequences connected with vertices, as well a datastructure for at most 4 out edges, for 'A', 'C', 'G' and 'T'. The edge stores integer number, how many times the edge is used in output eulerian path.}
  \label{fig:debruijn}
  \includegraphics[width=0.7\textwidth]{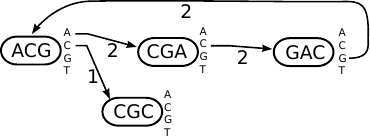}
\end{figure}

Moreover, the edges of the de Bruijn graph have an additional property, the integer number,
which depicts a number of occurrences of DNA fragments of length $k$ in the input set of DNA reads,
as in A-Bruijn graph~\cite{pevzner2004novo}.
It is worth noting that any vertex can have only up to 4 outgoing edges because we work over a 4-letter alphabet \{A, C, G, T\}.
The example of $4$ dimension graph ($k=4$) for $ACGACGACGC$ is depicted in~Fig.~\ref{fig:debruijn}.

Because of memory usage constraints many algorithms are using compacted de Bruijn graphs. A compacted de-Bruijn graph merges all maximal non-branching paths into single vertices. This allows to improve memory usage.

Parallel algorithms can use current multi-core processors more effectively. It is not easy to propose parallel algorithm,
because the independent threads of execution must communicate with each other in some way, which introduces a possibility of race conditions that must be eliminated.
The easiest techniques to eliminate race conditions: locks, e.g. mutexes, limit scalability, therefore are not the better solution for huge data structures and a large number of parallel threads.

A more challenging approach in parallel computing than eliminating conflicts using locks is the lock-free approach, where data structures and algorithms allowing conflicts are developed. When such a phenomenon occurs, the operation is repeated. It works very well if the conflict (race condition) is rare. Lock-free algorithms do not reduce scalability.
This approach uses special processor instructions called atomic operations, for example, compare-and-swap.

Example of such program is Jellyfish, which counts k-mers using lock-free approach. It does not however create de-Bruijn graph.

In this article we present a lock-free A-Bruijn graph of which example is depicted in Fig.~\ref{fig:debruijn}.
We provide lock-free algorith to build such graph from set of reads and algorithm for searching eulerian paths, that represent assembly output, in such graph.

We implemented our algorithm as C++ library, using C++11 atomic, and tested it on two platforms:
laptop with 4-core processor and dual XEON processor server, where we can execute 16 threads in parallel.
We used reads from sequencing databases for \textit{E. Coli}, \textit{Yeast} and \textit{Homo Sepiens}.
Our results were compared to other tools and depict the better speed and better scalability than previous methods.

\section{Methods}

\subsection{Graph data structure}

\begin{figure}[!htb]
  \caption{Lock-free A-Bruijn data straucture (on the left). It is array of $N$ atomic pointers for Nodes. Node (on the right) represent a vertex. It is structure with string view (k-mer), four atomic 8-bit integer counter, four atomic pointers to represents edges in graph and one non-atomic pointer to create single linked list of Nodes with the same hash.}
  \label{fig:hashtable}
  \includegraphics[width=0.7\textwidth]{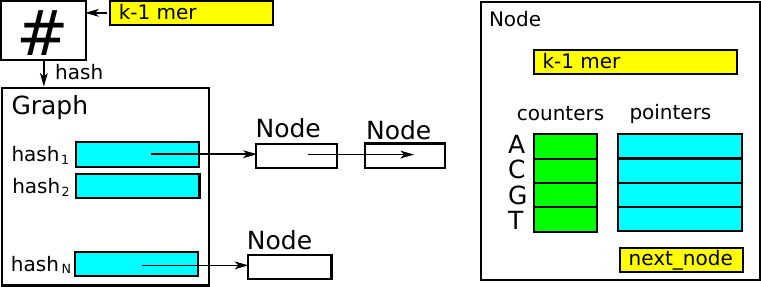}
\end{figure}

Graph is a~hash-table of verticles, and it is represented as array of $N$ atomic pointers to Nodes, as depicted in~Fig~\ref{fig:hashtable}.
Each element of array is a head of a single linked list of verticles, it is the most common implementation of hash tables.
Hash function converts the string connected with node into index $i$ with value from $0$ to $N-1$.

We assume the index store typically at most one node. In such case threads do not compete with each other in a sense of using lock-free structures.
We achieve it by both: (1) properly developed hash function that should uniformly spread available k-mers into values from $0$ to $N-1$,
(2) proper size of array that should be bigger than number of vertices.

For very rare cases, when we have different nodes with the same hash, the non-atomic single linked list of nodes is provided.
This case is also a reason why the string,  $k-1$ mer, is stored with the node.
This string is used to distinguish between vertices with the same hash value.
As depicted above, we reduce the usage of this single linked list providing proper size of array and hash function.
What is more, only head of this list should be atomic pointer, because other pointers are only read.

Hash function used in our algorithm is standard \texttt{std::hash}, available in C++ standard library. This implementation guaranties that for two different keys probability that they give the same hash are minimal, approaching \texttt{1.0/std::numeric\_limits<std::size\_t>\\::max()}. For 64 bit unsigned integer it is around $5.42\mathrm{e}{-20}$.

In our case this probability is bigger, because we limit hash numbers by the size of our hash table. It equals \(\frac{1}{N}\) where N is the size of hash table.

The Node object represents the vertex with its output edges. The members of Node are depicted in~Fig.~\ref{fig:hashtable}.
The output edges are stored in two 4-element tables:  the table with four atomic 8-bit integers for counters and the table with four atomic pointers to Node.
The size of table is 4 because we have at most four output edges (for 'A', 'C', 'G' and 'T').

\begin{figure}[!htb]
  \caption{Example of lock-free A-Bruijn graph. The data structures depicted in Fig.~\ref{fig:hashtable} represent graph depicted in Fig.~\ref{fig:debruijn}.
    Hash table size $N=10$, hash function results: $\#(ACG)=1$, $\#(CGC)=2$, $\#(CGA)=5$, $\#(GAC)=6$}.
  \label{fig:hashtable-example}
  \includegraphics[width=0.5\textwidth]{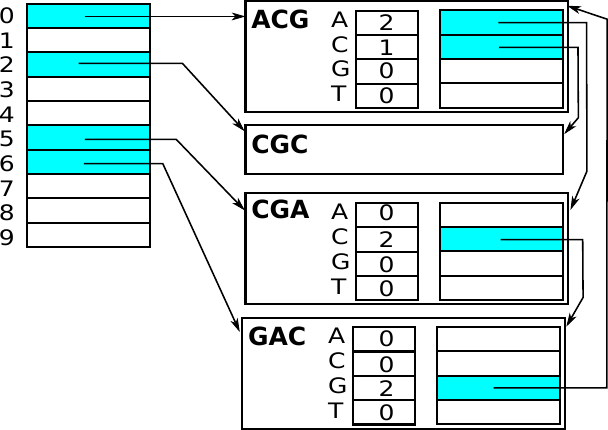}
\end{figure}

The example of lock-free graph represented by hash table with Nodes is presented in~Fig.~\ref{fig:hashtable-example}.

We use very fast method to convert label to counter index using ASCII representation of characters, only 2 assembler instructions:
\texttt{i = (c >{}>~1)~\&~0x3} where \texttt{>{}>} is right-site rotation of bites,
and \texttt{\&} is AND masking operation\footnote{Notation from C++ programming language}. Such mapping gives $i=0$ for label \texttt{'A'}, $i = 1$ for \texttt{'C'}, $i=2$ for \texttt{'T'}, and $i=3$ for \texttt{'G'}. The order of labels is different than depicted in~Fig.~\ref{fig:hashtable}, Fig.~\ref{fig:hashtable-example}. This is an implementation optimization and does not affect the algorithm.

8-bit counters means the we can represent maximum $255$ repetitions of motif that is longer than graph dimension. It is high enough in all genomes we worked on.

\subsection{Lock-Free graph creation algorithm}

The input of graph creation is large set of reads. Every thread in our program works on separate subset of reads.
From every read we obtain set of k-mers, substrings of length $k$ and this operation is performed locally inside threads, there is no races.
Then for every k-mer we add edge to the graph, with its source vertex and target vertex.
This operation should be provided in correct way.
When two or more processes are working on and changing the same data the races can occurs, therefore if not considered, obtained results are unpredictable.
To eliminate race conditions procedures have to be programmed using special methods.

First method uses mechanism that guarantee only one thread can be present in critical section, while others have to wait outside of it. Simple realization of this idea is called semaphore. It uses shared binary variable that is changed from 0 to 1 when one thread goes into critical section and from 1 to 0 when it goes out of it. When some other thread tries to enter critical section when semaphore is up, it has to wait for it to go down before going in itself. Such binary variable has to be free from race conditions itself. Such guarantee can give us atomic types. Operations modifying atomic types are indivisible by processor.

Second method does not use locks to limit number of threads in critical section. In order to avoid race conditions atomic instructions are used. While in first method this type of operations are used to create critical sections, here they are used directly on shared resources. This method eliminates the problem of race conditions and at the same time it does not block threads unnecessarily, which leads to better scalability in regards to active threads.

One of the main atomic function made available to developers in many programming languages is called Compare and Swap or CAS for short. CAS compares value in memory with one we expect and only if its the same changes it to the new value that we want. It then indicates whenever substitution ended in success or failure, because other thread already changed the value. Implementation of this function is used by our algorithm to insert vertices to the graph.
\begin{algorithm}[!htb]
\begin{algorithmic}[1]
\Function {CAS}{$val$,$ expected$,$ new\_val$}
\State $old\_val \gets val$
\If{$old\_val = expected$}
\State $reg \gets new\_val$
\EndIf
\State \Return $old\_val$
\EndFunction
\end{algorithmic}
\caption{Compare and Swap Function}
\label{alg:CAS}
\end{algorithm}

Lock-free method compared to locks guaranties system-wide progress, meaning that in some reasonable time at least one thread can make progress.
Our lock-free algorithm to build graph, called \texttt{MakeGraph}, is presented in~Alg.~\ref{alg:makegraph}.
This algorithm is run by every thread simultaneously. Please note that every thread has its own set of reads.
For every k-mer left and right vertices are created and added to the graph structure in lock-free way.
Then we create edge, connection between those vertices, which will be used when traversing the graph.

\begin{algorithm}[!htb]
\begin{algorithmic}[1]
\Function {makeGraph}{$reads$}
\ForAll{$r$ in $reads$}
\ForAll{$kmer$ in $r$}
\State $left \gets s_{0}s_{1}\dots{}s_{k-2}$ //prefix
\State $right \gets s_{1}\dots{}s_{k-2}s_{k-1}$ //suffix
\State $addLeftVertex(left, s_{k-1})$
\State $addRightVertex(right)$
\State {// now both left and right vertex exist}
\State $left\_vertex \gets FindVertex(left)$
\State $right\_vertex \gets FindVertex(right)$
\State $left\_vertex.pointers[s_{k-1}] \gets right\_vertex$
\EndFor
\EndFor
\EndFunction
\end{algorithmic}
\caption{Graph build algorithm}
\label{alg:makegraph}
\end{algorithm}

The~Alg~\ref{alg:addleftvertex} depicts adding left vertex.
We first have to search if it is already present in our graph. When we do not find it in lines from 2 to 6, we try to add it ourselves. While this thread was in the middle of going through the linked list other thread could have finished adding new vertex at the start of it. This is why we have to use CAS function. We compare first vertex of list with one we atomically read at line 8. If its the same it means that no other vertex was added and we can swap head with our new vertex. Otherwise our expected vertex (\texttt{vertex.next}) is updated to current head. Because head was changed by other thread it is possible that new vertex is one we want to add ourselves. Because of this before our next attempt at changing head we have to one more time go through our linked list in search of our vertex.

\begin{algorithm}[!htb]
\begin{algorithmic}[1]
\Function {addLeftVertex}{$left$, $next\_char$}
\If{$found = FindVertex(left)$}
\State $found.counters[next\_char]$ += 1 //add edge
\State \Return
\EndIf
\State $v \gets Vertex(left$) //not found, trying to add
\State $hash\gets \#(left)$
\State $v.next \gets AtomicLoad(table[hash]$)
\State $v.counters[next\_char]$ += 1 //add edge
\Do
\State $f \gets $ {\bf False}
\If{$found = FindVertex(left)$}
\State $found.counters[next\_char]$ += 1 //add edge
\State free(v) //remove temporary vertex
\State $f \gets $ {\bf True}
\EndIf
\doWhile { {\bf not} $f$ {\bf and not} $CAS(table[hash], v.next, v)$}
\EndFunction
\end{algorithmic}
\caption{Left vertex adding}
\label{alg:addleftvertex}
\end{algorithm}

Function addEdge increments proper weight of left vertex based on $next\_char$ which is last symbol of kmer. This allows us to count how many repeats of kmers we have in our reads.

Adding right vertex differs only in one thing from adding its left counterpart. When right vertex is already present in the structure we do not increment its edge. In fact we do not have to do anything else than to verify that it is present in our graph.

Find vertex function, presented in~Alg.~\ref{alg:findvertex}, searches for vertex with specific k-1-mer.
It does it by iterating through single linked list with vertices of particular hash value and returns either pointer to vertex or null pointer.

\begin{algorithm}[!htb]
\begin{algorithmic}[1]
\Function{findVertex}{$s$}
\State $hash\gets \#(s)$
\State $v \gets$ AtomicLoad($table[hash]$)
\While{$v$ {\bf not} $NULL$}
\If{$v.kmer = s$}
\State \Return $v$
\EndIf
\State $v \gets v.next$
\EndWhile
\State \Return $NULL$
\EndFunction
\end{algorithmic}
\caption{Finding vertex}
\label{alg:findvertex}
\end{algorithm}

\subsection{Normalization of graph}
After graph is created we have to normalize the weights of vertices. We do it by dividing every weight by the number of kmer duplicates we obtain in this calculation:
\[
\frac{number\_of\_reads * (read\_length -k +1)}{genome\_size}
\]
Every thread gets part of a hash table that it works on.
\subsection{Finding contigs}

When finding contigs we use a Boolean type variable called $first$ that is added to nodes. This variable depicts if the graph has an edge pointing to the vertex represented by the node. When the left vertex is created by Alg.~\ref{alg:addleftvertex}, its $first$ element is set to \textbf{true}. At that point, the graph has no edges pointed to the new vertex. However, $first$ of the right vertex is always set to \textbf{false}, even when node was already present in the graph.
\begin{algorithm}[!htb]
\begin{algorithmic}[1]
\Function{findContigs}{$start, end$}
\While{$start$ {\bf not} $end$}
\State $v \gets table[start]$
\If{$v = NULL$}
\State $continue$
\EndIf
\While{$v$ {\bf not} $NULL$}
\If{$node.first $ $OR$ $node.onlyOnePath() = false$}
\State $makePath(v)$
\EndIf
\State $v \gets v.next$
\EndWhile
\State $start \gets start + 1$
\EndWhile
\EndFunction
\end{algorithmic}
\caption{Finding contigs}
\label{alg:findcontigs}
\end{algorithm}
Every thread gets its part of the hash table and searches for vertices with either set $first$ member as \textbf{true}, or vertices with more than one outgoing edge.
If it finds node with $first$ set to true it creates contig following edges for every such vertex as long as only one edge goes out of the currently considered vertex.
If it finds other type of node it starts to create one contig for every such outgoing edge.

In both of there situations building of contig continues as long as there is only one outgoing edge of node currently being processed.

\subsection{Implementation issues}

When building graph a large number of nodes are allocated in a very short time. Such extensive use of heap can degrade performance when using multiple threads. This state of affairs can manifest because of memory allocator, which in many systems uses critical sections to handle requests, and thus handles requests sequentially. This can cause a lot of conflicts and threads waiting relatively long for the release of lock.

Object pools are structures that remedy this problem. They give us more control over how memory is managed. Most importantly, they can allocate sizeable chunks of memory for later use. When thread needs memory for allocation of vertex it gets it directly from pool object. In the program we use Object Pool structure available in boost library.

\subsection{Code organization}

The algorithm is developed in C++ version 20. It is a single class, \textbf{\texttt{LFdBG}} (Lock-Free-de-Bruijn-Graph), dependent on the standard C++ library and Boost.Pool\cite{schaling2011boost}. The \texttt{LFdBG.h} and \texttt{LFdBG.cpp} files together have approximately 370 lines of code.

When creating the \texttt{LFdBG} object, the user must provide a few fundamental parameters and easily available parameters: the graph dimension (lenght of k-mers, $k$), the number of threads to be used for assembly, the size of the hash table and approximate genome size.
The algorithm could be treated as parameterless for default values: $k=51$, number of threads equal $4 \times$ number of available cores, hash table equal genome size.

\section{Results}

We evaluated the performance of the LFdBG algorithm and compared them to the state-of-the-art algorithm Cuttlefish2\cite{10.1093/bioinformatics/btab309}.
We conducted experiments on server 4 $\times$ Xeon CPU E7-4830 v3 \@ 2.10GHz.
In testing we used sequences of:
\textit{Escherichia coli 536}  with $4,938,920$ bp ($NC\_008253$ in GenBank\cite{benson2000genbank}),
\textit{Saccharomyces cerevisiae S288C} (baker's yeast) with $12,157,105$ bp ($NC\_001133.9$\cite{benson2000genbank}),
\textit{Homo sapiens, GRCh38.p14} (human) chromosome 1 with $248,956,422$ bp ($NC\_000001.1$\cite{benson2000genbank}).

To measure the quality of the obtained contigs, we use the QUAST\cite{gurevich2013quast}. The individual metrics used in the research are:
\begin{itemize}
\item N50: length for which the set of contigs not shorter than it covers at least half of the reconstructed genome during assembly,
\item NG50: length for which the set of contigs not shorter than it covers at least half of the reference genome,
\item L50 (LG50): number of contigs equal to or greater in length than N50 (NG50).
\end{itemize}
QUAST considers only contigs whose lengths are at least 500 symbols.

We obtained the measurements of both algorithms on the same machine with the same input data.
For time measurement we used C++ standard \texttt{chrono} library\cite{josuttis2012c++} with \texttt{high\_resolution\_clock}. Since time for each thread for cpu-time sums up, we opted to measure wall-clock time. Time taken to load reads into memory by LFdBG algorithm is also taken into account.

Memory usage was measured using Unix \texttt{ps}(process status) command\cite{raymond2003art}.

\subsection{Data preparation}

Data preparation step includes:
\begin{itemize}
\item from \textit{E.~Coli} genome we created randomly chosen $200$ bp reads with coverage $=10\times$, $=30\times$ and $=50\times$.
Graph dimension ($k$) was set to $31$ for both: LFdBG and CuttleFish algorithms, table size for LFdBG algorithm was set to $5,000,000$;
\item from yeast genome we created randomly chosen $200$ bp reads with coverage $=10\times$, $=30\times$ and $=50\times$.
  $K = 31$ (as previously), table size = $20,000,000$;
\item from human first chromosome we created randomly chosen $200$ bp reads with coverage $=10\times$, $=30\times$ and $=50\times$.
  $K = 31$, table size = $400,000,000$.
\end{itemize}

\begin{figure}[!htb]
  \caption{Calculation time on multi-processor server.}
  \label{fig:Badanie1.1}
  \includegraphics[width=0.80\textwidth]{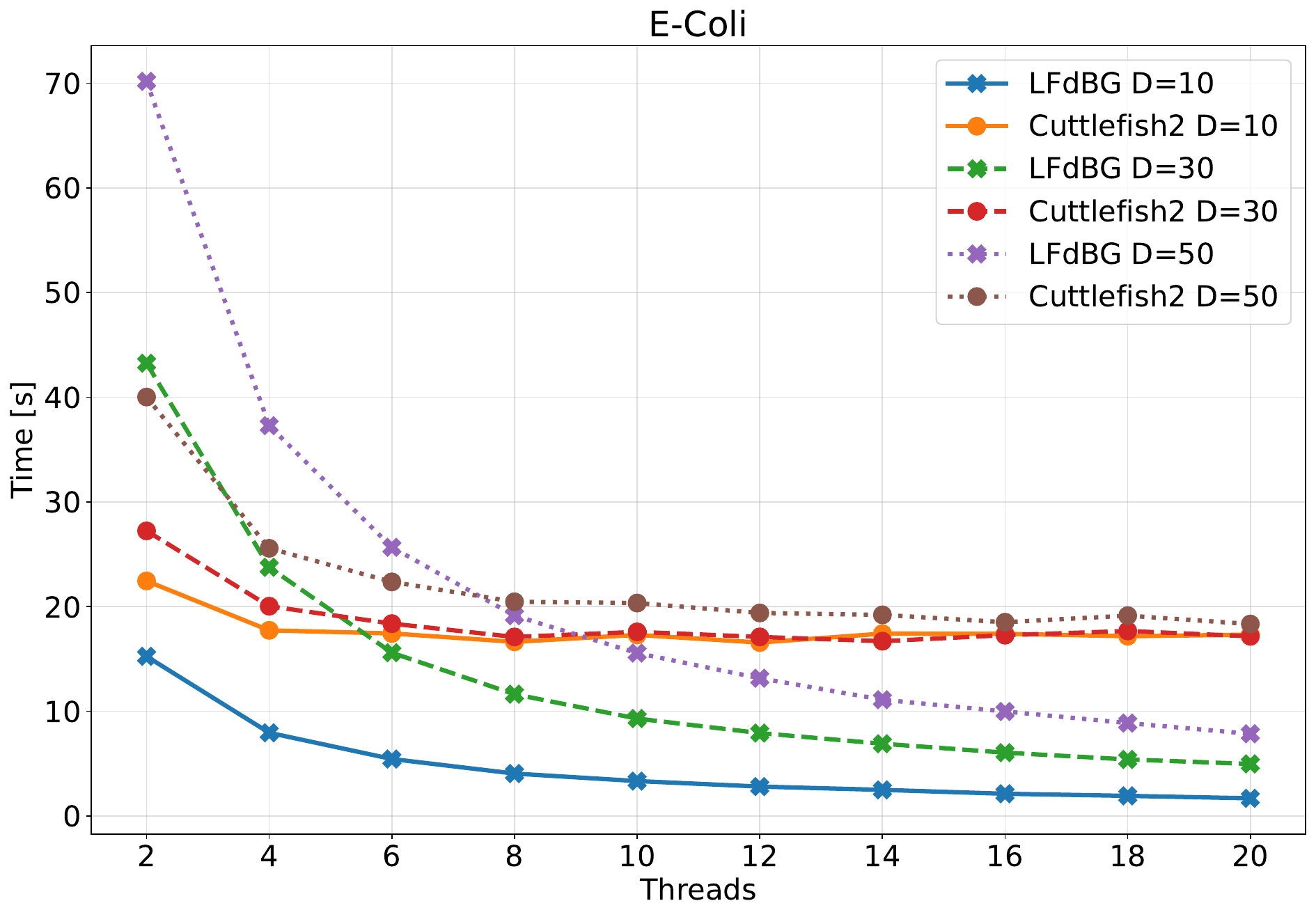}
\end{figure}

\begin{figure}[!htb]
  \caption{Calculation time on multi-processor server.}
  \label{fig:Badanie1.2}
  \includegraphics[width=0.80\textwidth]{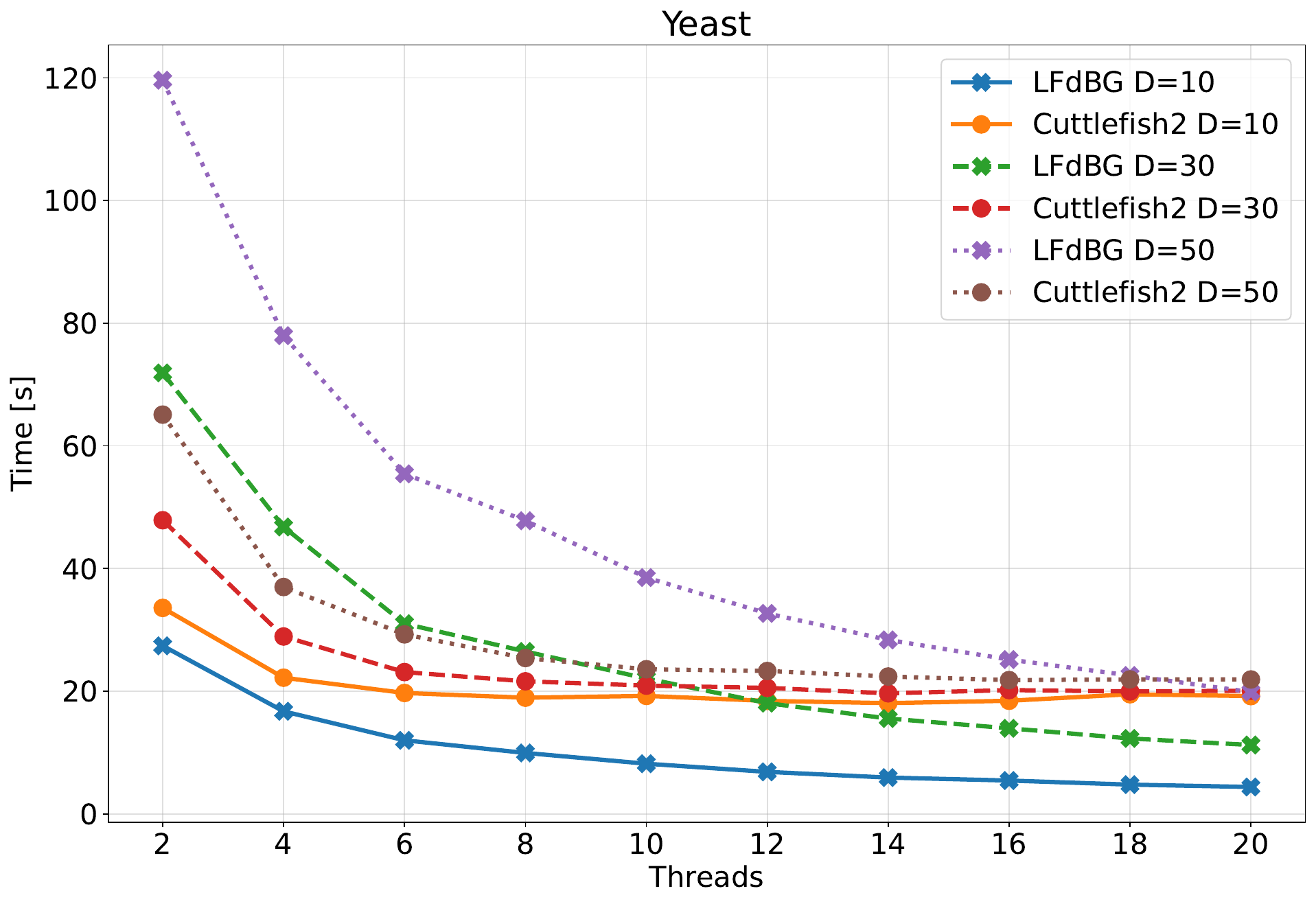}
\end{figure}

\begin{figure}[!htb]
  \caption{Calculation time on multi-processor server.}
  \label{fig:Badanie1.3}
  \includegraphics[width=0.80\textwidth]{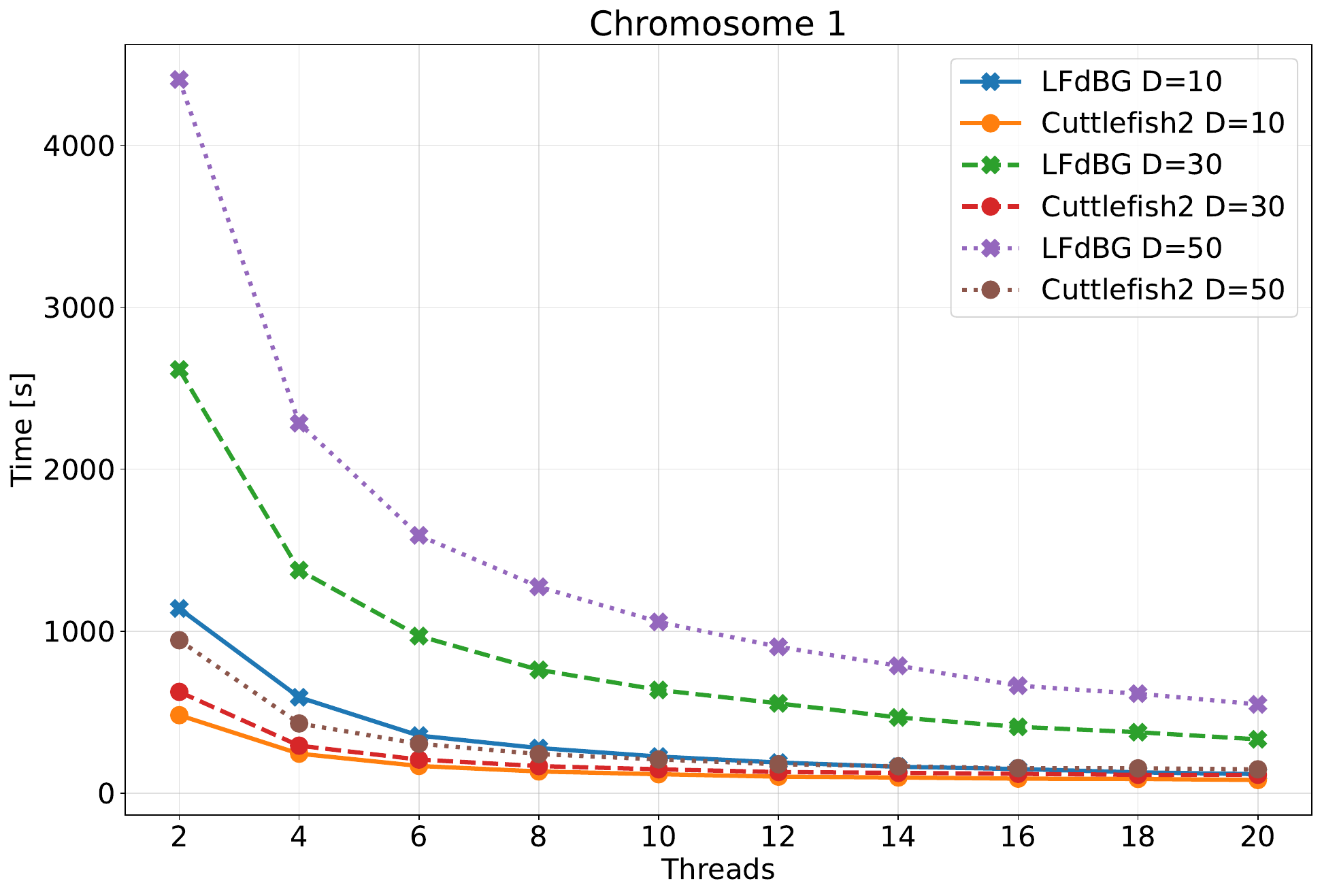}
\end{figure}

Results are depicted in Fig.~\ref{fig:Badanie1.1},  Fig.~\ref{fig:Badanie1.2} and  Fig.~\ref{fig:Badanie1.3}.

\subsection{Comparison of time performance and scalability}

LFdBG alogorithm is faster for all experiments with E-Coli and Yeast genomes where number of threads is large. When it comes to larger genome (Chromosome 1) then unfortunately LFdBG is slower. Both LFdBG and Cuttlefish2 are faster when there are more threads. What is more the speed up of LFdBG algorithm in regards of number of threads is better compared to Cuttlefish2.

\subsection{The impact of sequencing coverage on performance}

The greater the depth of coverage, the longer both algorithms run.
This is especially true with the LFdBG program, which does not filter repeated kmers in any way. On the other hand CuttleFish2 algorithm uses filters that can catch such duplicates. As a result, the operation time of this algorithm does not extend as much.

\subsection{Comparison of memory usage}

The memory usage for both algorithms is presented in Tab.~\ref{tab:memusage}.
CuttleFish2 has clear advantage in this experiments.

\begin{table}[!htb]
\caption{Used memory [GB]\label{tab:memusage}}
\centering
\begin{tabular}{|c|r|r|r|}
 \hline
 Genome & D & CF & LFdBG\\
 \hline
 E-Coli & 10 &0.10 & 0.52  \\
 \hline
  E-Coli & 30 &0.12 & 0.64  \\
 \hline
  E-Coli & 50 &0.21 & 0.76  \\
 \hline
 Baker's Yeast & 10 & 0.20 & 2.40 \\
 \hline
  Baker's Yeast & 30 & 0.32 & 2.68 \\
 \hline
  Baker's Yeast & 50 & 0.53 & 2.98 \\
 \hline
 Chromosome & 10 & 3.23 & 20.48 \\
 \hline
  Chromosome & 30 & 5.99 & 24.90 \\
 \hline
  Chromosome & 50 & 9.98 & 30.70 \\
 \hline
\end{tabular}
\end{table}

The faster rate at which LFdBG algorithm works is possible, because of a way that main structures are being processed.
LFdBG uses almost 6 times the memory compared to CuttleFish.
For example for working on first human choromosome with coverage 10x LFdBG needs $20.48$ GB RAM, compared to $3.23$ GB needed by CuttleFish.
However, for current servers, LFdBG is able to properly process whole human genome in $256$ GB RAM.

\subsection{The impact of read errors}
\begin{figure}[!htb]
  \includegraphics[width=0.80\textwidth]{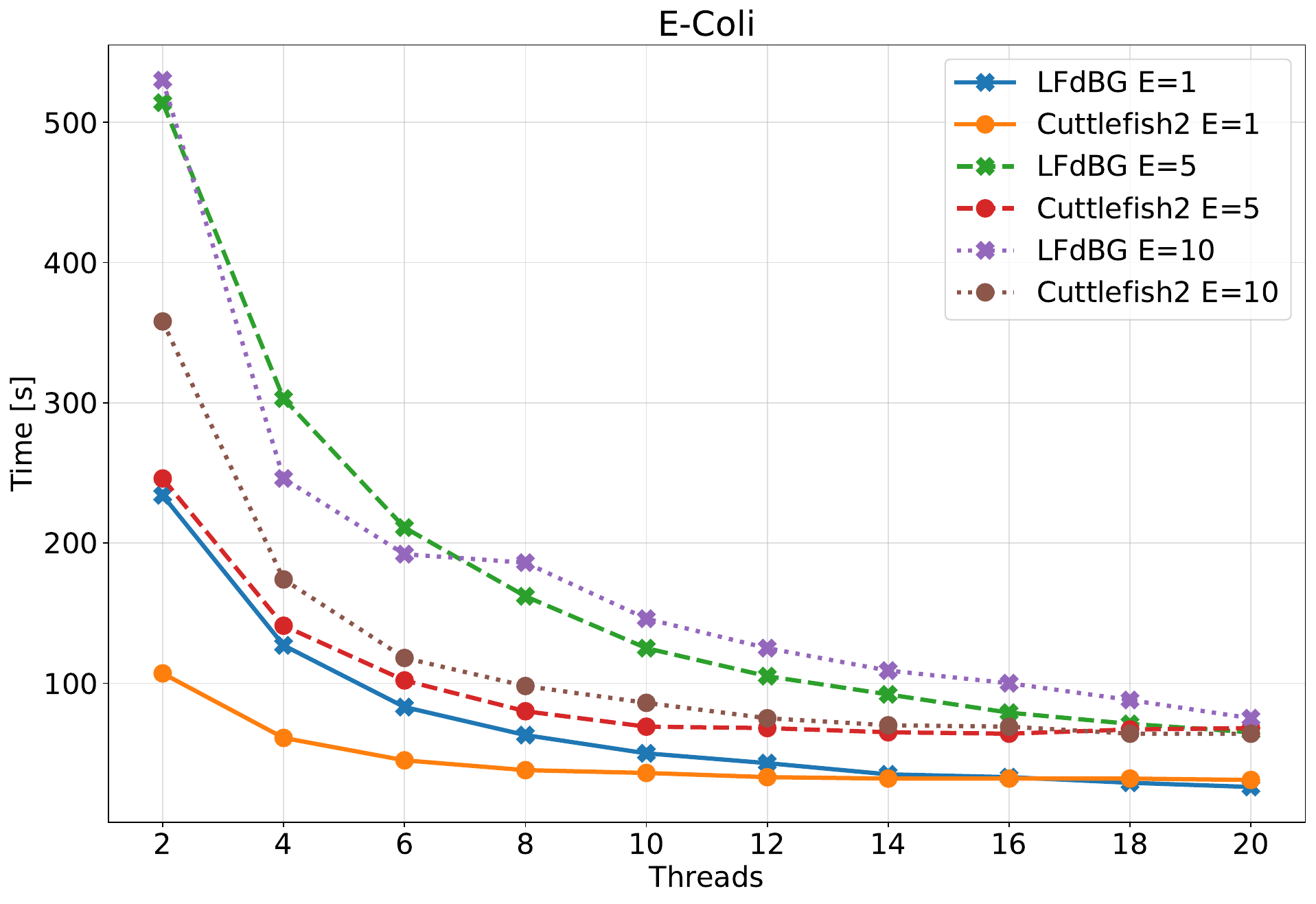}
  \caption{Calculation time on multi-processor server.}
\label{fig:Badanie1.5}
\end{figure}

\begin{figure}[!htb]
  \includegraphics[width=0.80\textwidth]{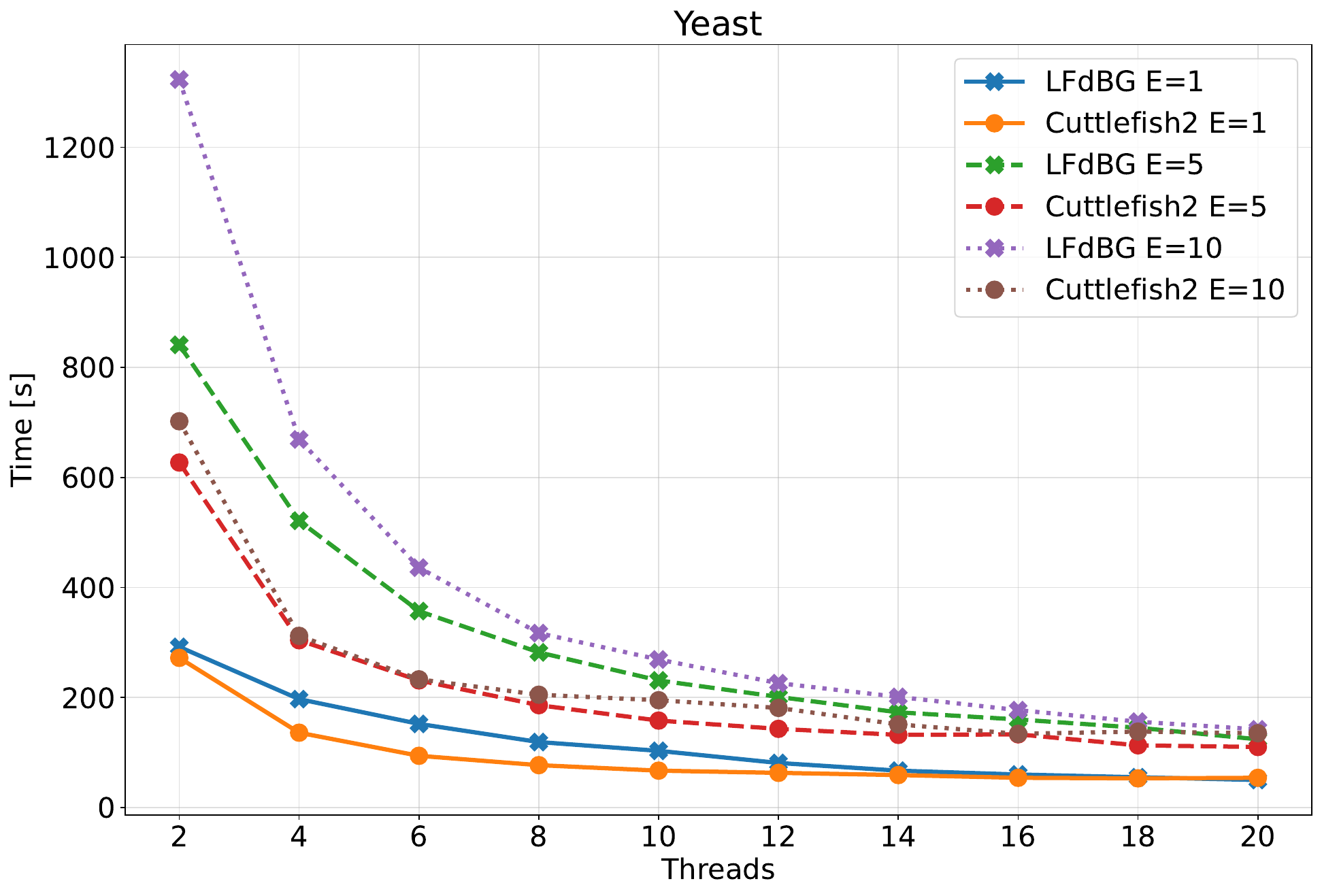}
  \caption{Calculation time on multi-processor server.}
\label{fig:Badanie1.6}
\end{figure}

The impact of read errors was measured using coverage 50 for E-Coli and Yeast genomes. We created 200bp reads with 1, 5, and 10\% error rate respectively. For both low and high error rate for small number of threads Cuttlefish2 is faster. Only at the last few threads number does LFdBG catch up to it. Results are depicted in Fig.~\ref{fig:Badanie1.5} and Fig.~\ref{fig:Badanie1.6}.

\subsection{Contigs quality}

Quality of the created contigs are comparable for both algorithms, which can be seen in table~\ref{tab:quality}.

\begin{table}[!htb]
\caption{QUAST metrices of assembly results\label{tab:quality}}
\centering
\begin{tabular}{|c|r|r|r|r|r|r|}
  \hline
  ~ & \multicolumn{2}{c|}{\textit{E.Coli}} & \multicolumn{2}{c|}{Yeast} & \multicolumn{2}{c|}{Human chr. 1}\\
  ~ & LFdBG & CF & LFdBG & CF & LFdBG & CF \\
 \hline
 contigs no.            & 263    & 300    & 887    & 987    & 83219 & 82218 \\
 longest contig [kbp]   & 129    & 129    & 147    & 108    & 26    & 32    \\
 genome assembly [\%]   & 97.60  & 96.35  & 93.66  & 92.20  & 71.79 & 70.27 \\
 N50   [kbp]            & 37  & 33  & 27  & 23  & 3  & 3  \\
 L50                    & 37     & 42     & 132    & 155    & 16687 & 16547 \\
 NG50  [kbp]            &37   & 31  & 26  & 21  & 2  & 2  \\
 LG50                   & 39     & 45     & 146    & 177    & 31192 & 32437 \\
 \hline
\end{tabular}
\end{table}

Despite the absence of errors in readings, both algorithms managed to assemble approximately only 70\% of the human chromosome as seen in table~\ref{tab:quality}. This is due to several factors. One of them is too small $K$, which means that from the point of view of the graph there were a lot of repetitions of substrings in the genome that it could not resolve. It should also be remembered that QUAST only takes into account contigs longer than 500 symbols.

\section{Conclusions}

This work aimed to create an algorithm for genome assembly using the lock-free structure of the de Bruijn graph. This algorithm was to be competitive with other available solutions and be easy to use.

The LFdBG algorithm is both fast and scales well with the increase in the number of its threads. The resulting contigs do not differ in quality from those created by the compared CuttleFish algorithm. Our algorithm is available as a C++ library that is easy to use in other programs.

\section{Availability of data and materials}%% if any

C++ library with source codes is available at github \url{https://github.com/dandon223/LFdBG} under MIT license.

\subsection{Funding}%% if any

This work was supported by Warsaw University of Technology grant CYBERIADA1.

\section{Authors' contributions}

DG and RN developed the solution, project the software and numerical experiments. DG implemented and tested software and performed numerical experiments under RN supervision. Both authors interpreted the results and prepared the manuscript. All authors read and approved the manuscript.

% if your bibliography is in bibtex format, use those commands:
\bibliographystyle{abbrv}
\bibliography{lfdb}

\begin{thebibliography}{1}

\bibitem{benson2000genbank}
D.~A. Benson, I.~Karsch-Mizrachi, D.~J. Lipman, J.~Ostell, B.~A. Rapp, and
  D.~L. Wheeler.
\newblock Genbank.
\newblock {\em Nucleic acids research}, 28(1):15--18, 2000.

\bibitem{gurevich2013quast}
A.~Gurevich, V.~Saveliev, N.~Vyahhi, and G.~Tesler.
\newblock Quast: quality assessment tool for genome assemblies.
\newblock {\em Bioinformatics}, 29(8):1072--1075, 2013.

\bibitem{josuttis2012c++}
N.~M. Josuttis.
\newblock The c++ standard library: a tutorial and reference.
\newblock 2012.

\bibitem{10.1093/bioinformatics/btab309}
J.~Khan and R.~Patro.
\newblock {Cuttlefish: fast, parallel and low-memory compaction of de Bruijn
  graphs from large-scale genome collections}.
\newblock {\em Bioinformatics}, 37(Supplement 1):i177--i186, 07 2021.

\bibitem{pevzner2004novo}
P.~A. Pevzner, H.~Tang, and G.~Tesler.
\newblock De novo repeat classification and fragment assembly.
\newblock {\em Genome research}, 14(9):1786--1796, 2004.

\bibitem{raymond2003art}
E.~S. Raymond.
\newblock {\em The art of Unix programming}.
\newblock Addison-Wesley Professional, 2003.

\bibitem{schaling2011boost}
B.~Sch{\"a}ling.
\newblock {\em The boost C++ libraries}.
\newblock Boris Sch{\"a}ling, 2011.

\end{thebibliography}

\end{document}